\documentclass[journal]{IEEEtran}
\ifCLASSINFOpdf
\else
\fi
\usepackage{etoolbox}
\usepackage[utf8]{inputenc} 
\usepackage[T1]{fontenc}    
\usepackage{hyperref}       
\usepackage{url}            
\usepackage{booktabs}       
\usepackage{amsfonts}       
\usepackage{nicefrac}       
\usepackage{microtype}      

\usepackage{times}
\usepackage{epsfig}
\usepackage{graphicx}
\usepackage{amsmath}
\usepackage{amssymb}
\usepackage{xcolor}
\usepackage{siunitx}
\usepackage{float}
\usepackage{booktabs}

\usepackage{bbm}
\usepackage{amsthm}
\usepackage{color, colortbl}
\usepackage{caption}
\usepackage{subcaption}

\definecolor{Gray}{gray}{0.9}
\definecolor{LightCyan}{rgb}{0.9,0.95,1}
\definecolor{LightRed}{rgb}{1,0.95,0.95}

\begin{document}
%
\title{SPICE: Self-supervised Pitch Estimation}
%
%
%

\author{Beat~Gfeller,
        Christian Frank,
        Dominik Roblek,
        Matt Sharifi,
        Marco Tagliasacchi,
        Mihajlo Velimirovi\'{c}
\thanks{All authors are with Google Research.}
\thanks{A shortened version of this manuscript is under review at ICASSP2020.}}

%
%

\markboth{Accepted to IEEE Trans. on Audio, Speech and Language Processing.}%
{Accepted to IEEE Trans. on Audio, Speech and Language Processing.}
%



\newcommand{\googlers}{\emph{SingingVoices}}
\newcommand{\mironek}{\emph{MIR-1k}}
\newcommand{\mdb}{\emph{MDB-stem-synth}}

\newcommand{\T}{T}
\newcommand{\Dim}{d}

\maketitle

\begin{abstract}
We propose a model to estimate the fundamental frequency in monophonic audio,
often referred to as pitch estimation. We acknowledge the fact that obtaining
ground truth annotations at the required temporal and frequency resolution is a
particularly daunting task. Therefore, we propose to adopt a self-supervised
learning technique, which is able to estimate pitch without any form
of supervision. The key observation is that pitch shift maps to a simple
translation when the audio signal is analysed through the lens of the constant-Q
transform (CQT). We design a self-supervised task by feeding two shifted slices
of the CQT to the same convolutional encoder, and require that the difference in
the outputs is proportional to the corresponding difference in pitch. In
addition, we introduce a small model head on top of the encoder, which is able
to determine the confidence of the pitch estimate, so as to distinguish between
voiced and unvoiced audio. Our results show that the proposed method is able to
estimate pitch at a level of accuracy comparable to fully supervised models,
both on clean and noisy audio samples, although it does not require access to large
labeled datasets.
\end{abstract}


\begin{figure*}[t]
  \centering
  \includegraphics[width=0.8\textwidth]{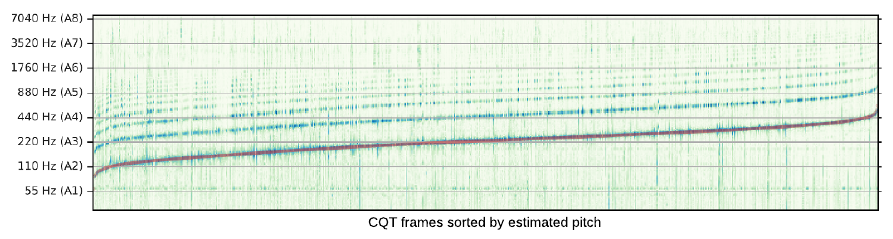}
  \caption{CQT frames extracted from the MIR-1k dataset re-ordered based on the pitch estimated by the SPICE algorithm (in red).}
  \label{fig:mir_onek}
\end{figure*}

%
\IEEEpeerreviewmaketitle

\section{Introduction}\label{sec:intro}

Pitch represents a perceptual property of sound which is \emph{relative}, since
it allows ordering to distinguish between high and low sounds, \emph{intensive},
that is, mixing sources with different pitches produces a chord, not a single
unified tone – contrary to loudness, which is additive in the number of sources,
and it is a property that can be attributed to a sound independently of its
\emph{source}~\cite{Lostanlen2016}. For example, the note A4 is perceived as the
same pitch whether it is played on a guitar or on a piano. A comprehensive
treatment of the psychoacoustic aspects of pitch perception is given
in~\cite{Lyon2017}. Pitch often corresponds to the fundamental frequency
($f_0$), i.e., the frequency of the lowest harmonic. However, the former is a
perceptual property, while the latter is a physical property of the underlying
audio signal. While there are a few notable exceptions (e.g., the Shepard tone,
the tritone paradox, or the auditory illusions described in~\cite{Esqueda2015}),
this correspondence holds for the broad class of locally periodic signals, which
represents a good abstraction for the audio signals considered in this paper.

Pitch estimation in monophonic audio received a great deal of attention over the
past decades, due to its central importance in several domains, ranging from
music information retrieval to speech analysis. Traditionally, simple signal
processing pipelines were proposed, working either in the time
domain~\cite{Dubnowski1976,Boersma1993,Talkin1995,DeCheveigne2002}, in the
frequency domain~\cite{Camacho2008} or both~\cite{Ramabadran2004,Kawahara2005},
often followed by post-processing algorithms to smooth the pitch
trajectories~\cite{Mauch2014, Salamon2012}. Until recently, machine learning
methods had not been able to outperform hand-crafted signal processing pipelines
targeting pitch estimation. This was due to the lack of annotated data, which is
particularly tedious and difficult to obtain at the temporal and frequency
resolution required to train fully supervised models. To overcome these
limitations, a synthetically generated dataset was proposed
in~\cite{Salamon2017}, obtained by re-synthesizing monophonic music tracks while
setting the fundamental frequency to the target ground truth. Using this
training data, the CREPE algorithm~\cite{Kim2018a} was able to achieve
state-of-the-art results when evaluated on the same dataset, outperforming
signal processing baselines, especially under noisy conditions.

In this paper we address the problem of lack of annotated data from a different
angle. Specifically, we rely on self-supervision, i.e., we define an auxiliary
task (also known as a \emph{pretext} task) which can be learned in a completely
unsupervised way. To devise this task, we started from the observation that for
humans, including professional musicians, it is typically much easier to
estimate relative pitch, related to the frequency interval between two notes,
than absolute pitch, related to the actual fundamental frequency~\cite{Ziv2014}.
Therefore, we design SPICE (Self-supervised PItCh Estimation) to solve a similar
task. More precisely, our network architecture consists of a convolutional
encoder which produces a single scalar embedding. We aim at learning a model
that linearly maps this scalar value to pitch, when the latter is expressed in a
logarithmic scale, i.e., in units of semitones of an equally tempered chromatic
scale. To do this, we feed two versions of the same signal to the encoder, one
being a pitch shifted version of the other by a random but known amount. Then,
we devise a loss function that forces the difference between the scalar
embeddings to be proportional to the known difference in pitch. Upon
convergence, the model is able to estimate relative pitch, solely relying on
self-supervision. In order to translate relative pitch to absolute pitch, we
apply a simple calibration step, which can be done using a small synthetically
generated dataset. Therefore, the model is able to produce absolute pitch
without having access to any manually labelled dataset.

A key characteristic of our model is that it receives as input a signal
transformed in the domain defined by the constant-Q transform (CQT), which
represents a convenient choice for analysing pitch. Indeed, the CQT filter bank
computes a wavelet transform~\cite{Anden2013}, and wavelets can be effectively
used to represent the class of locally periodic signals. When the number of
filters per octave (also known as quality factor) is large enough, wavelets have
a discernible pitch which is related to the logarithm of the scale variable. For
this reason, pitch shifting can be conveniently expressed as a simple
translation along the log-spaced frequency axis induced by the CQT. Note that
this property holds also for inharmonic or noisy audio signals for which the
fundamental frequency cannot be defined. For example, stretching these signals
in time produces a sensation of pitch shift, which would be observable in the
CQT domain despite the absence of the fundamental frequency. Conversely, we
acknowledge the fact that for some specific audio signals the analysis in the
CQT domain might lead to erroneous pitch estimates. For example, if the input
signal is a Shepard tone and the amount of pitch shift is equal to +11
semitones, the human ear would perceive a pitch interval of -1 semitone. Hence,
both the magnitude and the sign of the estimated pitch would be incorrect.
Despite the existence of these handcrafted examples for which our approach does
not apply, the correspondence between pitch shift and translation in the CQT
domain still holds for most real-world audio signals used to train and evaluate
our model.

Another important aspect of pitch estimation is determining whether the
underlying signal is voiced or unvoiced. Instead of relying on handcrafted
thresholding mechanisms, we augment the model in such a way that it can learn
the level of confidence of the pitch estimation. Namely, we add a simple fully
connected layer that receives as input the penultimate layer of the encoder and
produces a second scalar value which is trained to match the pitch estimation
error.

In summary, this paper makes the following key contributions:
\begin{itemize}
    \item We propose a self-supervised pitch estimation model, which can be
    trained without having access to any labelled dataset.
    \item We incorporate a self-supervised mechanism to estimate the confidence
    of the pitch estimation, which can be directly used for voicing detection.
    \item We evaluate our model against two publicly available monophonic
    datasets and show that in both cases we outperform handcrafted baselines,
    while matching the level of accuracy attained by CREPE, despite having no
    access to ground truth labels.
    \item We train and evaluate our model also in the noisy conditions, where
    background music is present in addition to monophonic singing, and show that
    also in this case, match the level of accuracy obtained by CREPE.
\end{itemize}

As an illustration, Figure~\ref{fig:mir_onek} shows the CQT frames of one of the
evaluation datasets (MIR-1k~\cite{Jang2009}), which are considered to be voiced.
The red solid line represents pitch as estimated by the SPICE and the CQT frames
are sorted according to this value from low to high pitch.

The rest of this paper is organized as follows. Section~\ref{sec:related_work}
contrasts the proposed method against the existing literature.
Section~\ref{sec:method} illustrates the proposed method, which is evaluated in
Section~\ref{sec:experiments}. Conclusions and future remarks are discussed in
Section~\ref{sec:conclusions}.

\section{Related work}\label{sec:related_work}
\textbf{Pitch estimation}: Traditional pitch estimation algorithms are based on
hand-crafted signal processing pipelines, working in the time and/or frequency
domain. The most common time-domain methods are based on the analysis of local
maxima of the auto-correlation function (ACF)~\cite{Dubnowski1976}. These
approaches are known to be prone to octave errors, because the peaks of the ACF
repeat at different lags. Therefore, several methods were introduced to be more
robust to such errors, including, e.g., the PRAAT~\cite{Boersma1993} and
RAPT~\cite{Talkin1995} algorithms. An alternative approach is pursued by the YIN
algorithm~\cite{DeCheveigne2002}, which looks for the local minima of the
Normalized Mean Difference Function (NMDF), to avoid octave errors caused by
signal amplitude changes. Different frequency-domain methods were also proposed,
based, e.g., on spectral peak picking~\cite{Martin1982} or template matching
with the spectrum of a sawtooth waveform~\cite{Camacho2008}. Other approaches
combine both time-domain and frequency-domain processing, like the Aurora
algorithm~\cite{Ramabadran2004} and the nearly defect-free F0 estimation
algorithm~\cite{Kawahara2005}. Comparative analyses including most of the
aforementioned approaches have been conducted on
speech~\cite{Jouvet2017,Strombergsson2016}, singing voices~\cite{Babacan2013}
and musical instruments~\cite{VonDemKnesebeck2010}. Machine learning models for
pitch estimation in speech were proposed in~\cite{Han2014,Lee2012}. The method
in ~\cite{Han2014} first extracts hand-crafted spectral domain features, and
then adopts a neural network (either a multi-layer perceptron or a recurrent
neural network) to compute the estimated pitch. In~\cite{Lee2012} consensus of
other pitch trackers is used to get ground truth, and a multi-layer perceptron
classifier is trained on the principal components of the autocorrelations of
subbands from an auditory filterbank. More recently the CREPE~\cite{Kim2018a}
model was proposed, an end-to-end convolutional neural network which consumes
audio directly in the time domain. The network is trained in a fully supervised
fashion, minimizing the cross-entropy loss between the ground truth pitch
annotations and the output of the model. In our experiments, we compare our
results with CREPE, which is the current state-of-the-art.

\textbf{Pitch confidence estimation}: Most of the aforementioned methods also
provide a voiced/unvoiced decision, often based on heuristic thresholds applied
to hand-crafted features. However, the confidence of the estimated pitch in the
voiced case is seldom provided. A few exceptions are CREPE~\cite{Kim2018a},
which produces a confidence score computed from the activations of the last
layer of the model, and~\cite{Deng2017}, which directly addresses this problem,
by training a neural network based on hand-crafted features to estimate the
confidence of the estimated pitch. In contrast, in our work we explicitly
augment the proposed model with a head aimed at estimating confidence in a fully
unsupervised way.

\textbf{Pitch tracking and polyphonic audio}: Often, post-processing is applied
to raw pitch estimates to smoothly track pitch contours over time. For example,
~\cite{Bonninghoff2016} applies Kalman filtering to smooth the output of a
hybrid spectro-temporal autocorrelation method, while the pYIN
algorithm~\cite{Mauch2014} builds on top of YIN, by applying Viterbi decoding of
a sequence soft pitch candidates. A similar smoothing algorithm is also used in
the publicly released version of CREPE~\cite{Kim2018a}. Pitch extraction in the
case of polyphonic audio remains an open research problem~\cite{Salamon2014}. In
this case, pitch tracking is even more important to be able to distinguish the
different melody lines~\cite{Salamon2012}. A machine learning model targeting
the estimation of multiple fundamental frequencies, melody, vocal and bass line
was recently proposed in~\cite{Bittner2018} .

\textbf{Self-supervised learning}: The widespread success of fully supervised
models was stimulated by the availability of annotated datasets. In those cases
in which labels are scarse or simply not available, self-supervised learning has
emerged as a promising approach for pre-training deep convolutional networks
both for vision~\cite{Noroozi2016,Wei2018,VandenOord2019} and audio-related
tasks~\cite{Jansen2018, Tagliasacchi2019, Meyer2017}. Somewhat related to our
paper are those methods that try to use self-supervision to obtain point
disparities between pairs of images~\cite{Christiansen2019}, where shifts in the
spatial domain play the role of shifts in the \mbox{log-frequency} domain.

\newcommand{\fbase}{f_{base}}
\newcommand{\x}{\mathbf{x}}
\newcommand{\xh}{\hat{\x}}
\newcommand{\Enc}{Enc}
\newcommand{\Dec}{Dec}
\newcommand{\Loss}{\mathcal{L}}
\newcommand{\Losspitch}{\mathcal{L}_{\text{pitch}}}
\newcommand{\Lossrecon}{\mathcal{L}_{\text{recon}}}
\newcommand{\Lossconf}{\mathcal{L}_{\text{conf}}}
\newcommand{\pitchscale}{\sigma}
\newcommand{\Fmax}{F_{\text{max}}}

\begin{figure*}[t]
    \centering
    \includegraphics[width=0.8\textwidth]{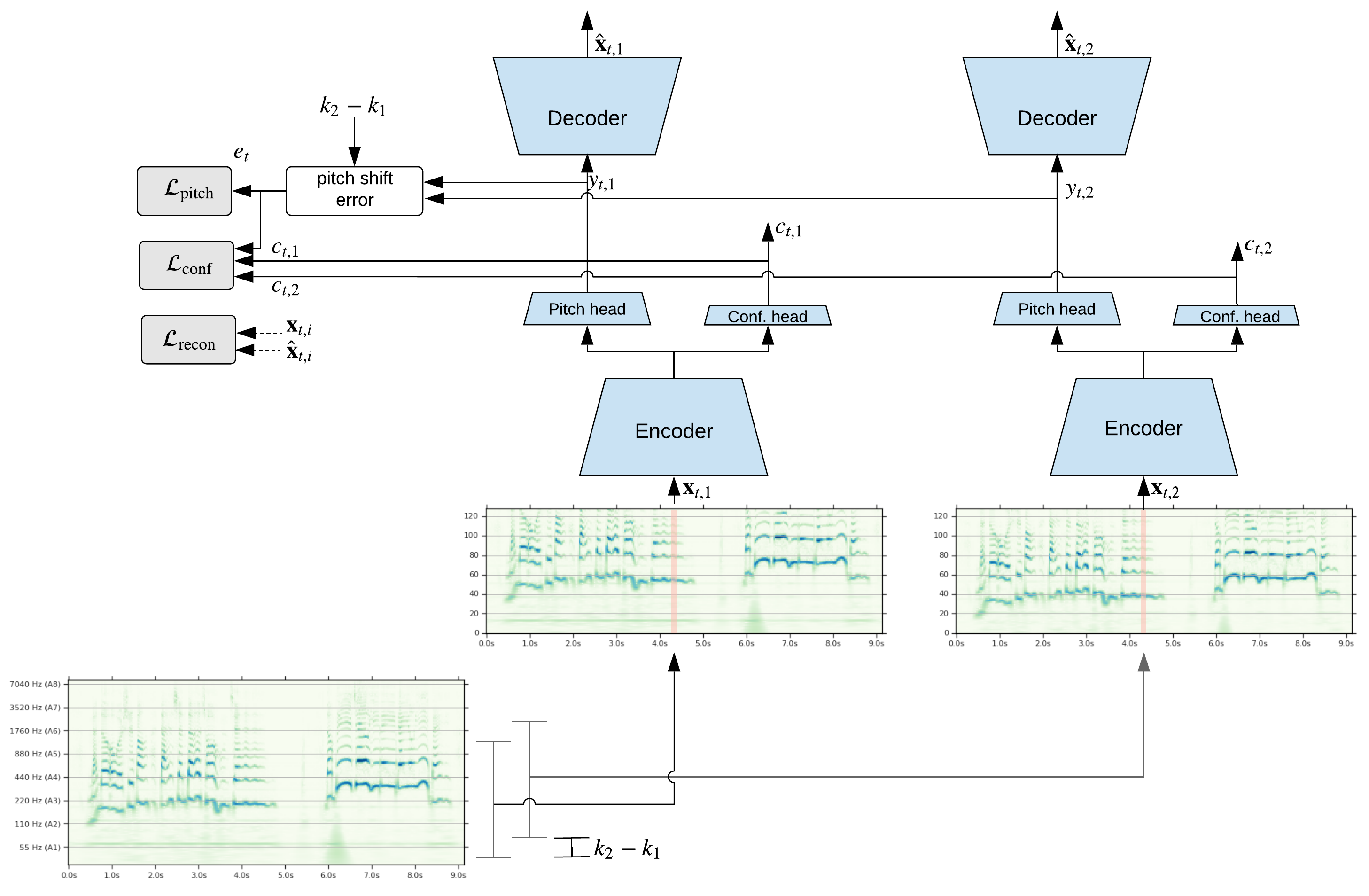}
    \caption{SPICE model architecture.}
    \label{fig:model_architecture}
\end{figure*}

\section{Methods}\label{sec:method}

The proposed pitch estimation model receives as input an audio track of
arbitrary length and produces as output a time series of estimated pitch
frequencies, together with an indication of the confidence of the estimates. The
latter is used to discriminate between unvoiced frames, in which pitch is not
well defined, and voiced frames.

\subsection*{Audio frontend}
Our proposed model does not consume audio directly, but instead it receives as
input individual frames of the constant-Q transform (CQT). As illustrated
in~\cite{Anden2013}, the CQT representation approximately corresponds to the
output of a wavelet filter bank defined by the following family of wavelets:
\begin{equation}
  \psi_{\lambda_k}(t) = \lambda_k \psi( \lambda_k t) ,
\end{equation}
where $Q$ denotes the number of filters per octave and
\begin{equation}
   \lambda_k = \fbase 2 ^ {\frac{k}{Q}}, \quad k \in 0, \ldots, \Fmax - 1,
\end{equation}
where $\fbase$ is the frequency of the lowest frequency bin and $\Fmax$ is the
number of CQT bins. The Fourier transform of the wavelet filters can be
expressed as:
\begin{equation}
  \Psi_{\lambda_k} = \Psi\left(\frac{f}{\lambda_k}\right)
\end{equation}
Assuming that the center frequency of $\Psi(f)$ is normalized to 1, each filter
is centered at frequency $\lambda_k$ and has a bandwidth equal to $\lambda_k /
Q$. Hence, if we consider two filters with indices $k_1$ and $k_2$, one of the
corresponding wavelets would the pitch-shifted version of the other. That is,
\begin{equation}
  \Delta k = k_2 - k_1 = Q \cdot \log_2 \alpha,
\end{equation}
where $\alpha = \lambda_{k_2} / \lambda_{k_1}$.
Therefore, for the class of locally periodic signals that can be represented as
a wavelet expansion, a translation of $\Delta k$ bins in the CQT domain is
related to a pitch-shift by a factor $\alpha$.

Note that this key property of mapping pitch-shift to a simple translation  does
not hold in general for other audio frontends, e.g., for the widely used mel
spectrogram. In this case, the relationship between frequency (in Hz) and mel
units is given by
\begin{equation}
  m = c \cdot \log \left (1 + \frac{f}{f_{\text{break}}}\right )
\end{equation}
for some constants $c$ and $f_{\text{break}}$ (also known as break frequency).
Hence, the relationship is approximately linear at low frequencies ($f \ll
f_{\text{break}}$) and logarithmic at high frequencies ($f \gg
f_{\text{break}}$), with a smooth transition between these two regimes. It is
straightforward to show that a multiplicative scaling of frequencies does not
correspond to an additive scaling in the mel domain.

\subsection*{Pitch estimation}
The proposed model architecture is illustrated in
Figure~\ref{fig:model_architecture}. Given an input track, the audio frontend
computes the absolute value of the CQT, which is represented as a real-valued
matrix $X$ of size $T \times \Fmax$, where $T$ depends on the selected hop
length. From each temporal frame $t = 1,\ldots, T$ (where $T$ is equal to the
batch size during training) the model samples at random two integer offsets
$k_{t,1}$ and $k_{t,2}$ from a uniform distribution, i.e., $k_{t,i} \sim
\mathcal{U}(k_{\text{min}}, k_{\text{max}}$), and it extracts two corresponding
slices $\x_{t,1}, \x_{t,2} \in \mathbb{R}^F$, spanning the range of CQT bins
$[k_{t,i}, k_{t,i} + F]$, $i =1, 2$, where $F$ is the number of CQT bins in the
slice. Then, each vector is fed to the same encoder to produce a single scalar
$y_{t,i} = \Enc(x_{t,i}) \in \mathbb{R}$. The encoder is a neural network with
$L$ convolutional layers followed by two fully-connected layers. Further details
about the model architecture are provided in Section~\ref{sec:experiments}.

We design our main loss in such a way that $y_{t,i}$ is encouraged to encode
pitch. First, we define the relative pitch error as
\begin{equation}
    e_t = | (y_{t,1} - y_{t,2}) - \pitchscale (k_{t,1} - k_{t,2}) |
\end{equation}
Then, the loss is defined as the Huber norm~\cite{Huber1964} of the pitch error:
\begin{equation}
    \Losspitch = \frac{1}{T}\sum_t h(e_t),
\end{equation}
where:
\begin{equation}
    h(x) =
\begin{cases}
    \frac{\x^2}{2}, \quad |x| \le \tau \\
    \frac{\tau^2}{2} + \tau (| x | - \tau), \quad \text{otherwise.} \\
\end{cases}
\end{equation}

The pitch difference scaling factor $\pitchscale$ is adjusted in such a way that
$y_t \in [0, 1]$ when pitch is in the range $[f_{\text{min}}, f_{\text{max}}]$,
namely:
\begin{equation}
    \pitchscale = \frac{1}{Q \cdot \log_2(f_{\max} / f_{\min})}
\end{equation}
The values of $f_{\min}$ and $f_{\max}$ are determined based on the range of
pitch frequencies spanned by the training set. In our experiments we found that
the Huber loss makes the model less sensitive to the presence of unvoiced frames
in the training dataset, for which the relative pitch error can be large, as
pitch is not well defined in this case.

In addition to $\Losspitch$, we also use the following reconstruction loss
\begin{equation}
    \Lossrecon = \frac{1}{T}\sum_t \| \x_{t, 1} - \xh_{t, 1}\|_2^2 + \| \x_{t, 2} - \xh_{t, 2}\|_2^2,
\end{equation}

where $\xh_{t,i}$, $i=1,2$, is a reconstruction of the input frame obtained by
feeding $y_{i,t}$ into a decoder $\xh_{t,i} = \Dec(y_{i,t})$.
The reconstruction loss $\Lossrecon$ forces the reconstructed frame $\xh_{t,i}$
to be as close as possible to the original frame $\x_{t,i}$. The decoder is a
neural network with $L$ convolutional layers whose architecture is the mirrored
version of the encoder, with convolutions replaced by transposed convolutions,
which maps the scalar value $y_{i,t}$ back to a vector with the same shape as
the input frame. Further details about the model architecture are provided in
Section~\ref{sec:experiments}. In Section~\ref{sec:experiments} we also
empirically evaluate the impact of this loss component as part of the ablation
study.

Therefore, the overall loss is defined as:
\begin{equation}\label{eq:total_loss}
    \Loss = w_{\text{pitch}}\Losspitch + w_{\text{recon}}\Lossrecon,
\end{equation}
where $w_{\text{pitch}}$ and $w_{\text{recon}}$ are scalar weights that
determine the relative importance assigned to the two loss components.

Given the way it is designed, the proposed model can only estimate relative
pitch differences. The absolute pitch of an input frame is obtained by applying
an affine mapping:
\begin{equation}\label{eq:calibration}
    \hat{p}_{0,t} = b + s \cdot y_t = b + s \cdot \Enc(\x_t) \quad [\text{semitones}],
\end{equation}
which depends on two parameters. This is needed to map the output of the encoder
$y_t$ from the $[0, 1]$ range to the absolute pitch range (expressed in
semitones). We use a small amount of synthetically generated data (locally
periodic signals with a known frequency) to estimate both the intercept
$\hat{b}$ and the slope $\hat{s}$. More specifically, we generate a waveform
which is piece-wise harmonic and consists of $M$ pieces. Each piece is a purely
harmonic signal with fundamental frequency $f_0$ corresponding to a semitone
sampled uniformly at random in the range A2 (110Hz) and A4 (440Hz). We sample
the amplitude of the first harmonic in $a_0 \sim \mathcal{N}(0, 1)$ and that of
higher order harmonics in $a_k \sim a_0 \cdot \mathcal{U}(0, 1)$, $k=1, \ldots,
K$. A random phase is applied to each harmonic. In our experiments, each piece
is $N \cdot H$ samples long, where $H$ denotes the CQT hop-length used by the
SPICE model and $N$ the number of frames. We feed this waveform to SPICE and
consider the estimate produced for the central frame in each piece (to mitigate
errors due to boundary effects). This leads to $M$ synthetically generated
samples that can be used to fit the model in \eqref{eq:calibration}. In
Section~\ref{subsec:calibration} we empirically evaluate the robustness of the
calibration process for different values of $M$.

Note that pitch in \eqref{eq:calibration} is expressed in semitones and it can
be converted to frequency (in Hz) by:
\begin{equation}
  \hat{f}_{0,t} = \fbase 2^{\frac{\hat{p}_{0,t}}{12}}  \quad [\text{Hz}]
\end{equation}

\subsection*{Confidence estimation}
In addition to the estimated pitch $\hat{p}_{0,t}$, we design our model such
that it also produces a confidence level $c_t \in [0, 1]$. Indeed, when the
input audio is voiced we expect to produce high confidence estimates, while when
it is unvoiced pitch is not well defined and the output confidence should be
low. To achieve this, we design the encoder architecture to have two heads on
top of the convolutional layers, as illustrated in
Figure~\ref{fig:model_architecture}. The first head consists of two
fully-connected layers and produces the pitch estimate $y_t$. The second head
consists of a single fully-connected layer and produces the confidence level
$c_t$. To train the latter, we add the following loss:
\begin{equation}
    \Lossconf = \frac{1}{T}\sum_t | (1 - c_{t,1}) - e_t / \sigma |^2 + | (1 - c_{t,2}) - e_t / \sigma |^2.
\end{equation}

This way the model will produce high confidence $c_t \sim 1$ when the model is
able to correctly estimate the pitch difference between the two input slices. At
the same time, given that our primary goal is to accurately estimate pitch,
during the backpropagation step we stop the gradients so that $\Lossconf$ only
influences the training of the confidence head and does not affect the
other layers of the encoder architecture.

\begin{table*}[t]
  \centering
  \caption{Dataset specifications.}\label{tab:dataset_specs}
   \begin{tabular}{lcccccc}
      \hline
       &  & \multicolumn{3}{c}{Length} & \multicolumn{2}{c}{\# of frames} \\
      Dataset & \# of tracks &  min & max & total & voiced & total\\
      \hline
      {\mironek} & 1000 & 3s & 12s & 133m & 175k & 215k \\
      {\mdb} & 230 & 2s & 565s & 418m & 784k & 1.75M \\
      {\googlers} & 88 & 25s & 298s & 185m & 194k & 348k \\
      \hline
   \end{tabular}
\end{table*}

\begin{figure*}[t]
  \begin{subfigure}{0.32\textwidth}
    \centering
    \includegraphics[width=1.0\linewidth]{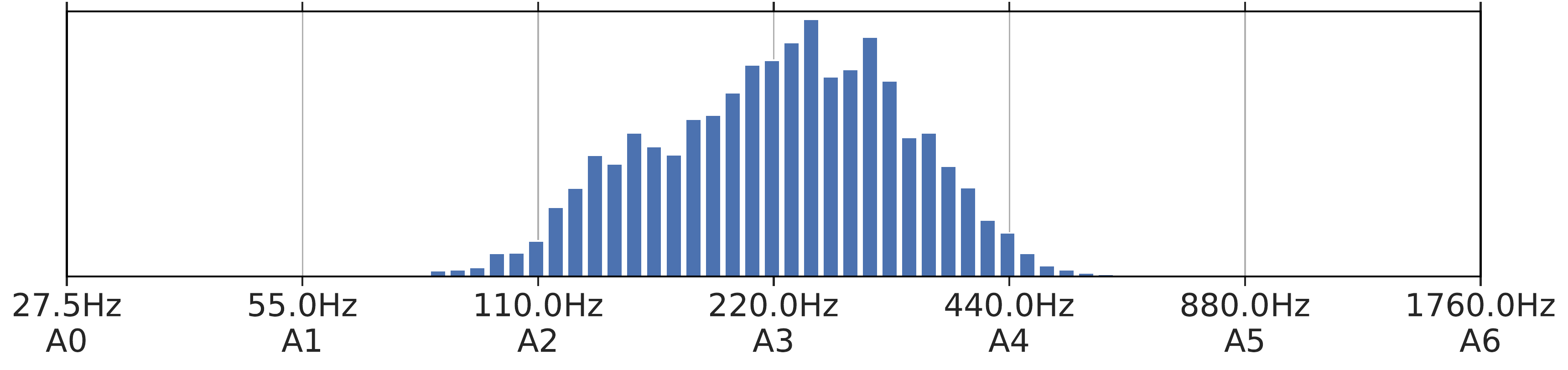}
    \caption{{\mironek}}
    \label{fig:hist_mironek}
  \end{subfigure}
  \begin{subfigure}{0.32\textwidth}
    \centering
    \includegraphics[width=1.0\linewidth]{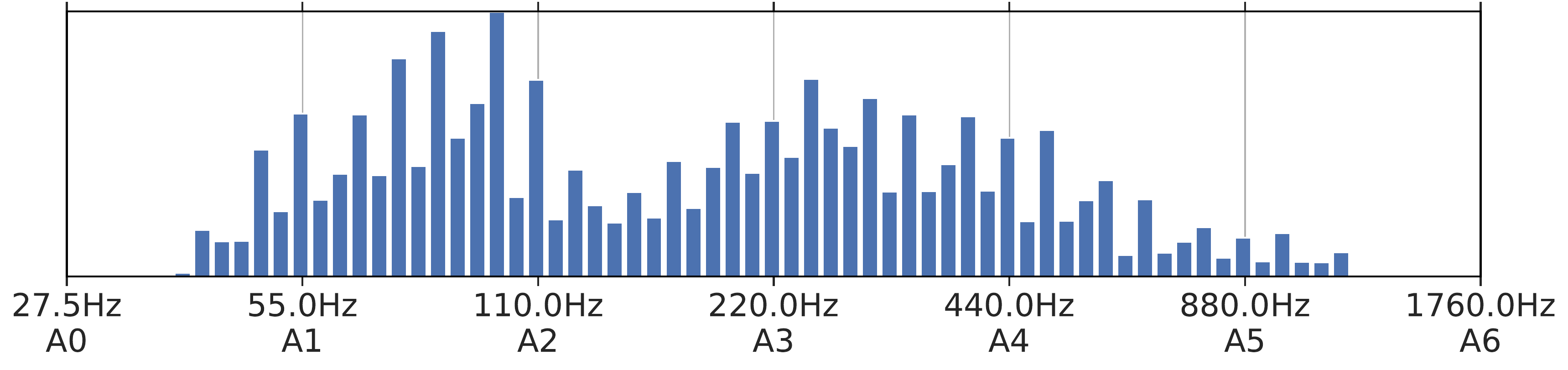}
    \caption{{\mdb}}
    \label{fig:hist_mdb}
  \end{subfigure}
  \begin{subfigure}{0.32\textwidth}
    \centering
    \includegraphics[width=1.0\linewidth]{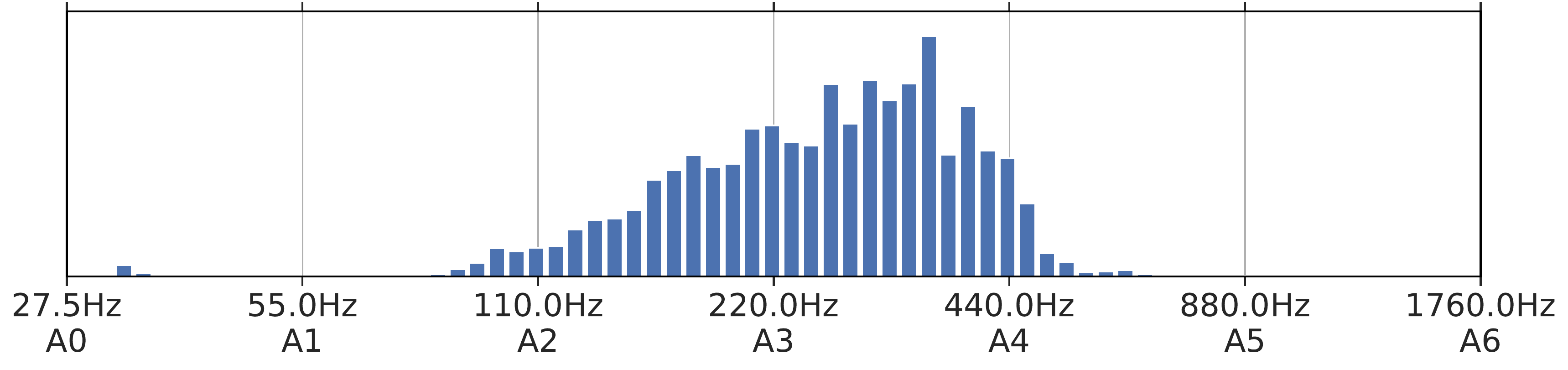}
    \caption{{\googlers}}
    \label{fig:hist_googlers}
  \end{subfigure}
  \caption{Range of pitch values covered by the different datasets.}
  \label{fig:dataset_pitch_hist}
  \end{figure*}

\subsection*{Handling background music}
\newcommand{\xc}{\x_{t,i}^{c}}
\newcommand{\xn}{\x_{t,i}^{n}}
\newcommand{\yc}{y_{t,i}^{c}}
\newcommand{\yn}{y_{t,i}^{n}}

The accuracy of pitch estimation can be severely affected when dealing with
noisy conditions, which emerge, for example, when the singing voice is
superimposed over background music. In this case, we are faced with polyphonic
audio and we want the model to focus only on the singing voice source. To deal
with these conditions, we introduce a data augmentation step in our training
setup. More specifically, we mix the clean singing voice signal with the
corresponding instrumental backing track at different levels of signal-to-noise
(SNR) ratios. Interestingly, we found that simply augmenting the training data
was not sufficient to achieve a good level of robustness. Instead, we also
modified the definition of the loss functions as follows. Let $\xc$ and $\xn$
denote, respectively, the CQT of the clean and noisy input samples. Similarly,
$\yc$ and $\yn$ denote the corresponding outputs of the encoder. The pitch error
loss is modified by averaging four different variants of the error, that is:

\begin{equation}
  e_t^{pq} = | (y_{t,1}^p - y_{t,2}^q) - \sigma(k_{t,1} - k_{t,2})|\quad p, q \in \{c, n\},
\end{equation}
\begin{equation}
  \Losspitch = \frac{1}{4}\sum_t \sum_{p,q \in \{c, n\}}  h(e_t^{pq}).
\end{equation}

The reconstruction loss is also modified, so that the decoder is asked to
reconstruct the clean samples only. That is:
\begin{equation}
  \Lossrecon = \frac{1}{T}\sum_t \| \x_{t, 1}^c - \xh_{t, 1}\|_2^2 + \| \x_{t, 2}^c - \xh_{t, 2}\|_2^2.
\end{equation}
The rationale behind this approach is that the encoder is induced to represent
in its output only the information relative to the clean input audio samples,
thus learning to denoise the input by separating the singing voice from noise.
\section{Experiments}\label{sec:experiments}

\subsection*{Model parameters}
First we provide the details of the default parameters used in our model. The
input audio track is sampled at $16$~kHz. The CQT frontend is parametrized to
use $Q = 24$ bins per octave, so as to achieve a resolution equal to one
half-semitone per bin. We set $\fbase$ equal to the frequency of the note $C_1$,
i.e., $\fbase \simeq 32.70$ Hz and we compute up to $F_{max} = 190$ CQT bins,
i.e., to cover the range of frequency up to Nyquist. We use a Hann window with
hop length set equal to 512 samples, i.e., one CQT frame every 32 ms. The CQT is
implemented using TensorFlow operations following the specifications of the
open-source \emph{librosa} library~\cite{Mcfee2015}. During training, we extract
slices of $F = 128$ CQT bins, setting $k_{\text{min}} = 0$ and $k_{\text{max}} =
8$ (i.e., between 0 and 4 semitones when $Q=24$). The Huber threshold is set to
$\tau = 0.25\pitchscale$ and the loss weights equal to, respectively,
$w_{\text{pitch}} = 10^4$ and $w_{\text{recon}} = 1$. We increased the weight of
the pitch-shift loss to $w_{\text{pitch}} = 3 \cdot 10^5$ when training with
background music.

The encoder receives as input a 128-dimensional vector corresponding to a
sliced CQT frame and produces as output two scalars representing, respectively,
pitch and confidence. The model architecture consists of $L = 6$ convolutional
layers. We use filters of size $3$ and stride equal to $1$. The number of
channels is equal to $d \cdot [1, 2, 4, 8, 8, 8]$, where $d = 64$ for the
encoder and $d = 32$ for the decoder. Each convolution is followed by batch
normalization and a ReLU non-linearity. Max-pooling of size $3$ and stride $2$
is applied at the output of each layer. Hence, after flattening the output of
the last convolutional layer we obtain an embedding of size $1024$ elements.
This is fed into two different heads. The pitch estimation head consists of two
fully-connected layers with, respectively, 48 and 1 units. The confidence head
consists of a single fully-connected layer with 1 output unit. The total number
of parameters of the encoder is equal to 2.38M. Note that we do not apply any
form of temporal smoothing to the output of the model.

The model is trained using Adam~\cite{Kingma2015} with default hyperparameters
and learning rate equal to $10^{-4}$. The batch size is set to $64$. During
training, the CQT frames of the input audio tracks are shuffled, so that the
frames in a batch are likely to come from different tracks.

\begin{figure}[t]
  \begin{subfigure}{0.49\textwidth}
    \centering
    \includegraphics[width=0.6\linewidth]{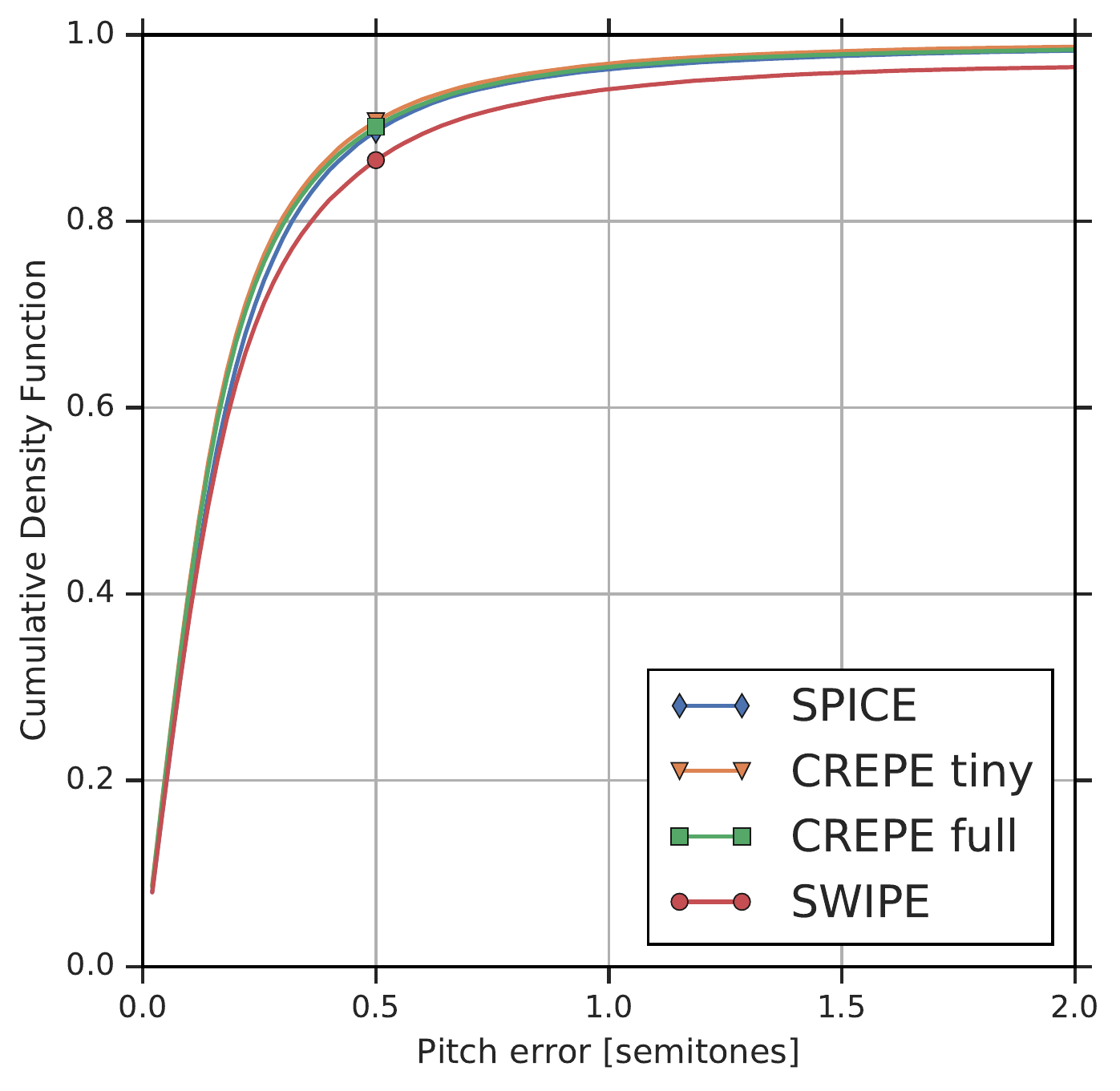}
    \caption{{\mironek}}
    \label{fig:rpa_mir_onek}
  \end{subfigure}
  \begin{subfigure}{0.49\textwidth}
    \centering
    \includegraphics[width=0.6\linewidth]{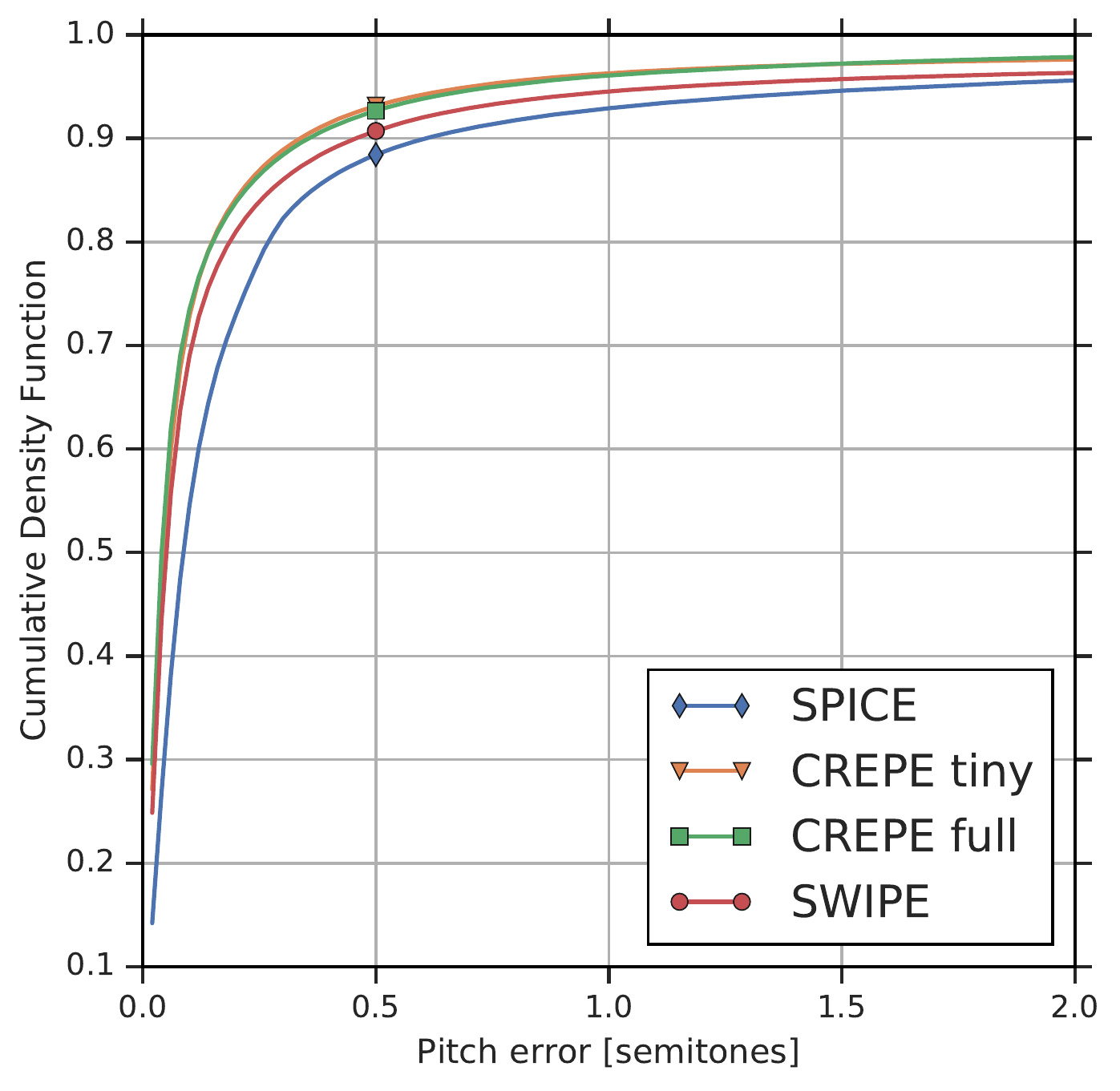}
    \caption{{\mdb}}
    \label{fig:vrr_mir_onek}
  \end{subfigure}
  \caption{Raw Pitch Accuracy.}
  \label{fig:evaluation_rpa}
\end{figure}

\begin{table*}[t]
  \centering
  \caption{Evaluation results.}\label{tab:main_eval}
   \begin{tabular}{lcccccc}
      \hline
       &  & & \multicolumn{2}{c}{{\mironek}} & \multicolumn{1}{c}{{\mdb}} \\
        Model & \# params & Trained on & RPA (CI 95\%) & VRR & RPA (CI 95\%) \\
        \hline
        SWIPE & - & - & 86.6\% & - & 90.7\% \\
        CREPE tiny & 487k &  many & 90.7\% & 88.9\% & 93.1\% \\
        CREPE full & 22.2M & many & 90.1\% & 84.6\% & 92.7\% \\
        SPICE & 2.38M & {\googlers} & $90.6\% \pm 0.1\%$ & 86.8\%
        & $89.1\% \pm 0.4\%$ \\
        SPICE & 180k & {\googlers} & $90.4\% \pm 0.1\%$ & 90.5\%
        & $87.9\% \pm 0.9\%$ \\
      \hline
   \end{tabular}
\end{table*}

\begin{table*}[t]
  \centering
  \caption{Evaluation results of the ablation study.}\label{tab:ablation_study}
  \begin{tabular}{lcccc}
    \hline
     & \multicolumn{1}{c}{{\mironek}} & \multicolumn{1}{c}{{\mdb}} \\
        &  RPA (CI 95\%) & RPA (CI 95\%) \\
      \hline
      SPICE baseline & $90.6\% \pm 0.1\%$
      & $89.1\% \pm 0.4\%$ \\
      \hline
       w/o reconstruction loss & $55.9\% \pm 2.3\%$
      & $56.3\% \pm 2.7\%$ \\
       w/o data augmentation & $89.6\% \pm 0.2\%$
      & $71.1\% \pm 0.7\%$ \\
       w/ continuous pitch shift augmentation & $90.7\% \pm 0.1\%$
      & $87.8\% \pm 0.8\%$ \\
       w/ L2 loss  & 89.2$\% \pm 0.2\%$
      & $86.3\% \pm 0.9\%$ \\
       w/ L1 loss  & $82.1\% \pm 0.7\%$
      & $82.3\% \pm 0.9\%$ \\
    \hline
 \end{tabular}
\end{table*}

\begin{figure*}[t]
  \begin{subfigure}{0.49\textwidth}
    \centering
    \includegraphics[width=1.0\linewidth]{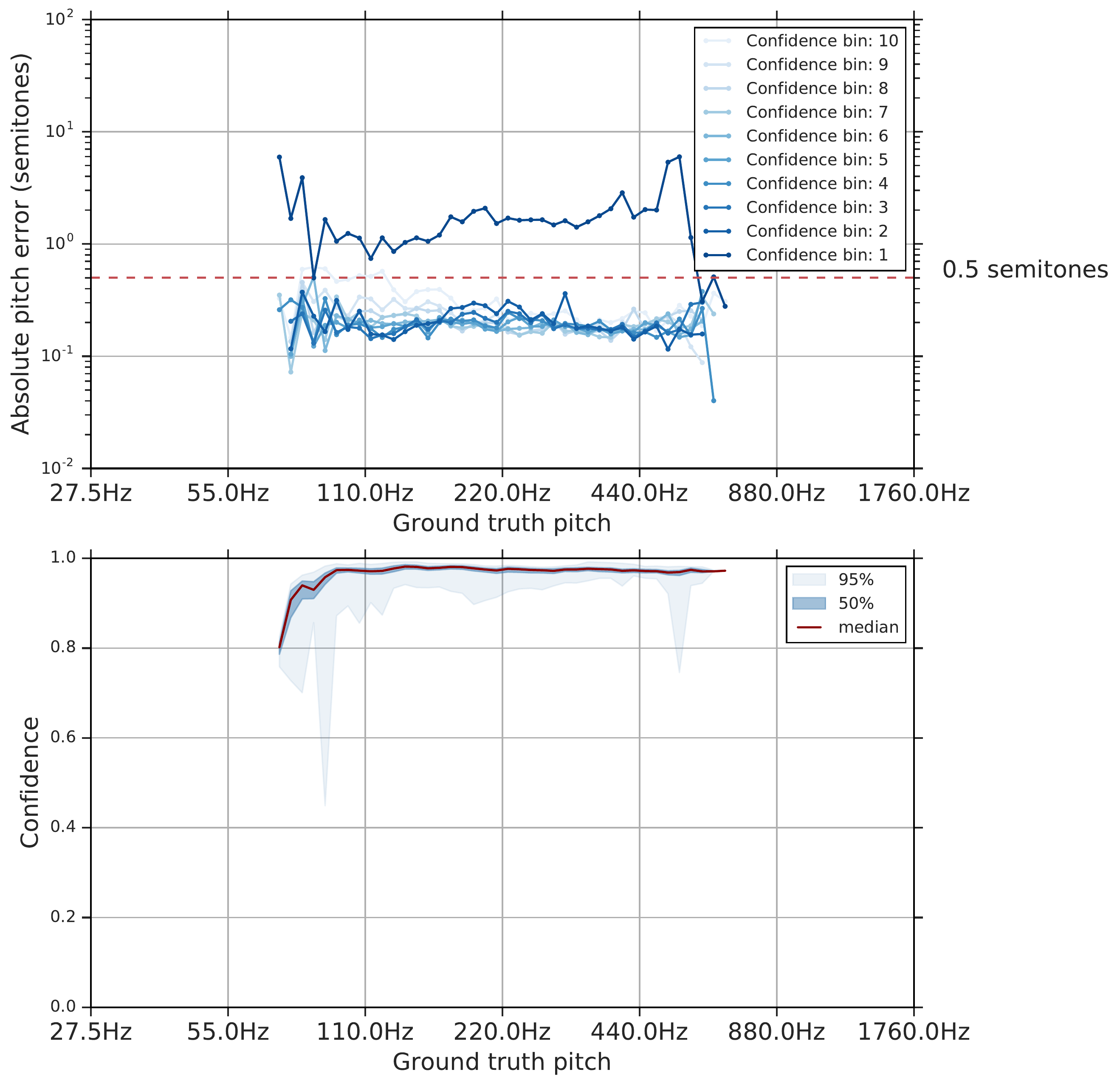}
    \caption{SPICE}
    \label{fig:error_vs_frequency_spice_mir_onek}
  \end{subfigure}
  \begin{subfigure}{0.49\textwidth}
    \centering
    \includegraphics[width=1.0\linewidth]{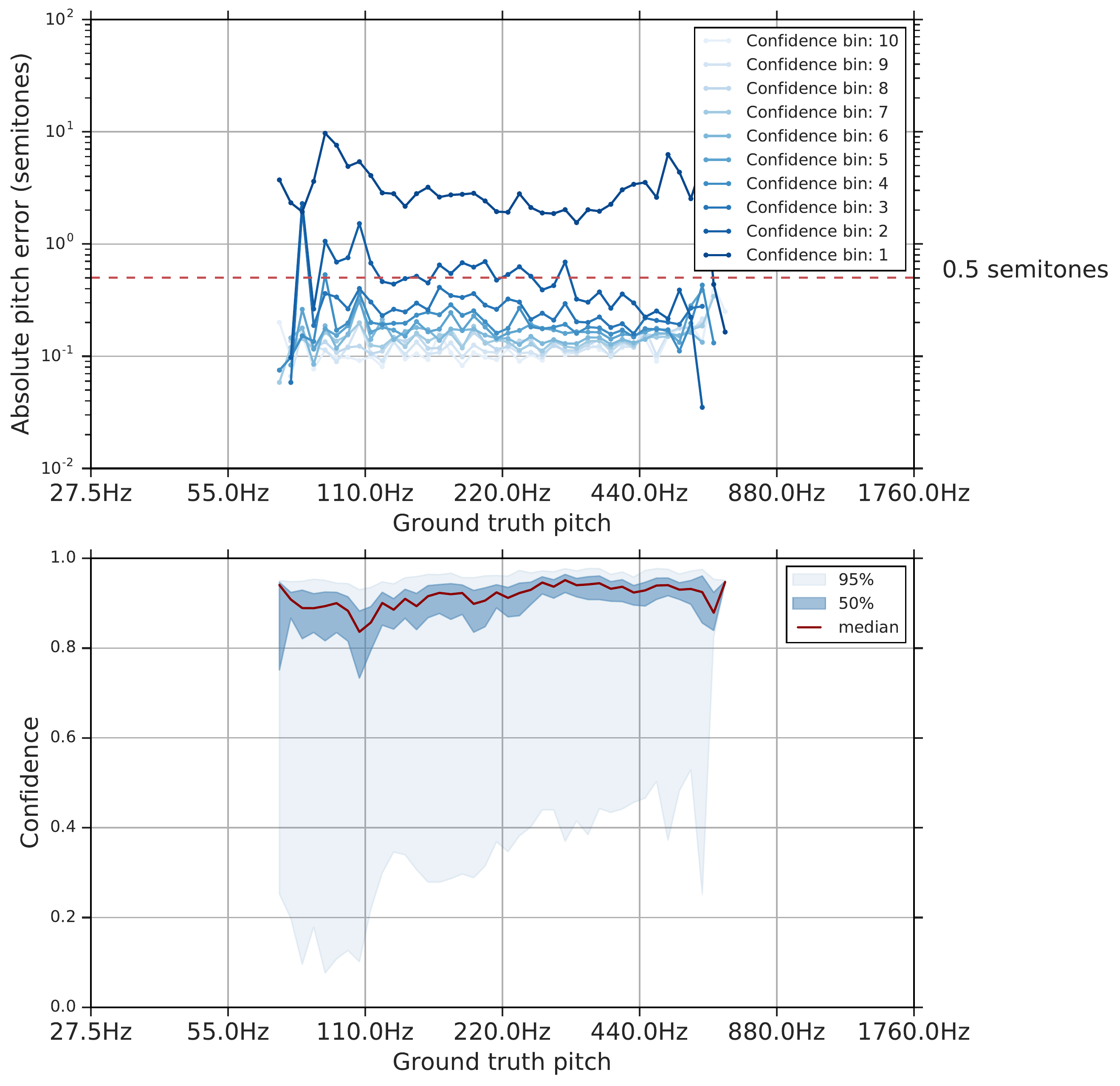}
    \caption{CREPE full}
    \label{fig:error_vs_frequency_crepe_full_mir_onek}
  \end{subfigure}
  \caption{Pitch error on the {\mironek} dataset, conditional on ground truth pitch and model confidence.}
  \label{fig:pitch_error_mir_onek}
\end{figure*}

\subsection*{Datasets}
We use three datasets in our experiments, whose details are summarized in
Table~\ref{tab:dataset_specs}. The {\mironek}~\cite{Jang2009} dataset contains
1000 audio tracks with people singing Chinese pop songs. The dataset is
annotated with pitch at a granularity of 10~ms and it also contains
voiced/unvoiced frame annotations. It comes with two stereo channels
representing, respectively, the singing voice and the accompaniment music. The
{\mdb} dataset~\cite{Salamon2017} includes re-synthesized monophonic music
played with a variety of musical instruments. This dataset was used to train the
CREPE model in~\cite{Kim2018a}. In this case, pitch annotations are available at
a granularity of 29~ms. Given the mismatch of the sampling period of the pitch
annotations across datasets, we resample the pitch time-series with a period
equal to the hop length of the CQT, i.e., 32~ms. In addition to these publicly
available datasets, we also collected in-house the {\googlers} dataset, which
contains 88 audio tracks of people singing a variety of pop songs, for a total
of 185 minutes.

Figure~\ref{fig:dataset_pitch_hist} illustrates the empirical distribution of
pitch values. For {\googlers}, there are no ground-truth pitch labels, so we
used the ouput of CREPE (configured with full model capacity and enabling
Viterbi smoothing) as a surrogate. We observe that {\mdb} spans a significantly
larger range of frequencies (approx. 5 octaves) than {\mironek} and {\googlers}
(approx. 3 octaves).
Note that this is still smaller than the range covered by human hearing, which
extends to 9-10 octaves. Further work is needed to collect datasets and
evaluate pitch estimation algorithms on such a broad frequency range.

We trained SPICE using either {\googlers} or {\mironek} and used both
{\mironek} (singing voice channel only) and {\mdb} to evaluate models
in clean conditions.
To handle background music, we repeated training on {\mironek}, but
this time applying data augmentation by mixing in backing tracks with a SNR
uniformly sampled from [-5dB, 25dB]. For the evaluation, we used the {\mironek}
dataset, mixing the available backing tracks at different levels of SNR, namely
20dB, 10dB and 0dB. In all cases, we apply data augmentation during training, by
pitch-shifting the input audio tracks by an amount in semitones uniformly
sampled in the discrete set $\{-12, 0, +12\}$,
using a TensorFlow-based implementation of the phase-vocoder algorithm
in~\cite{Laroche1999}. We found that this works better than sampling from a
continuous set, because of the artifacts introduced by the pitch-shifting
algorithm when adopting arbitrary rational scaling factors.

\begin{figure*}[t]
  \begin{subfigure}{0.49\textwidth}
    \centering
    \includegraphics[width=1.0\linewidth]{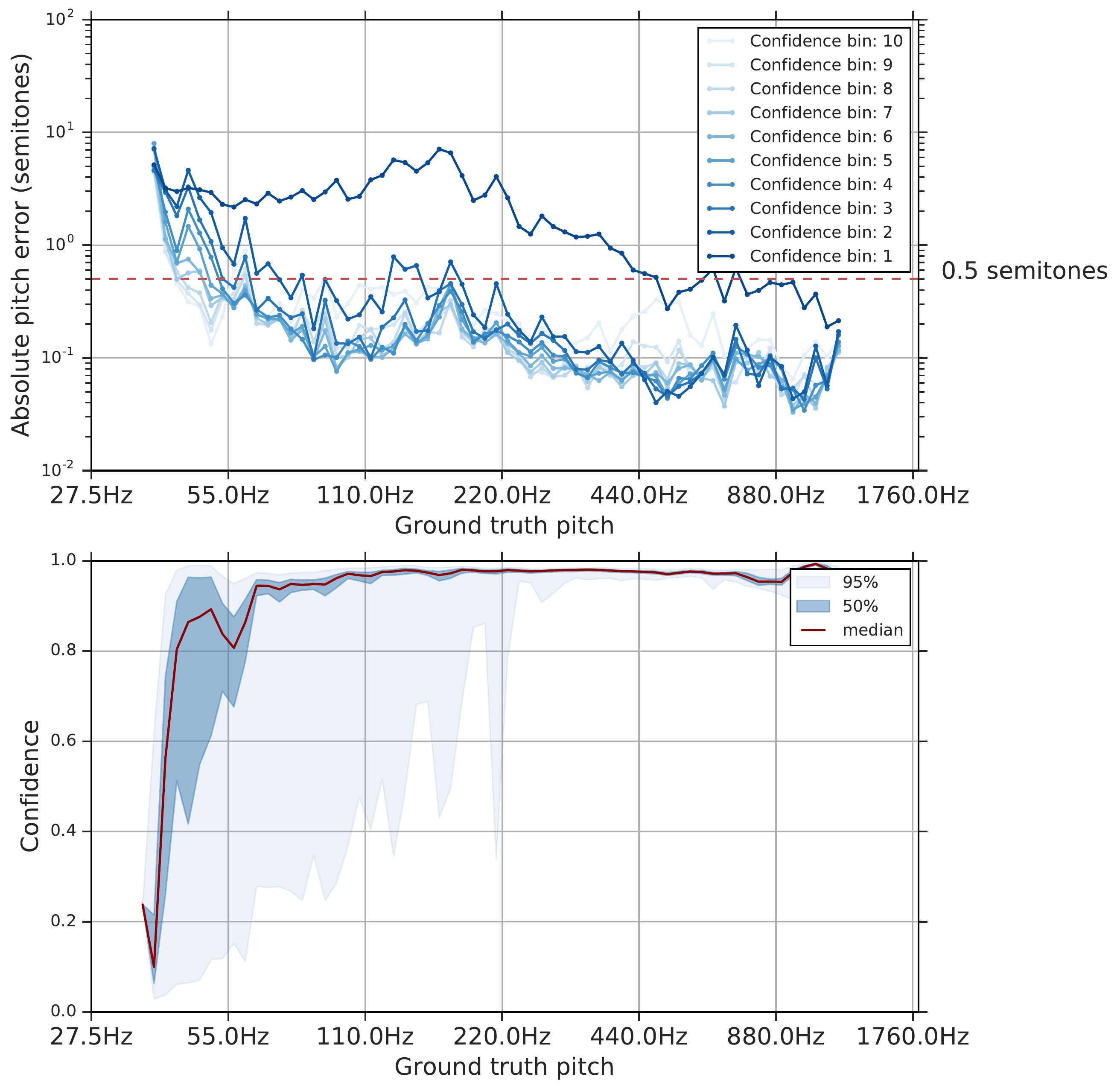}
    \caption{SPICE}
    \label{fig:error_vs_frequency_spice_mdb_stem_synth}
  \end{subfigure}
  \begin{subfigure}{0.49\textwidth}
    \centering
    \includegraphics[width=1.0\linewidth]{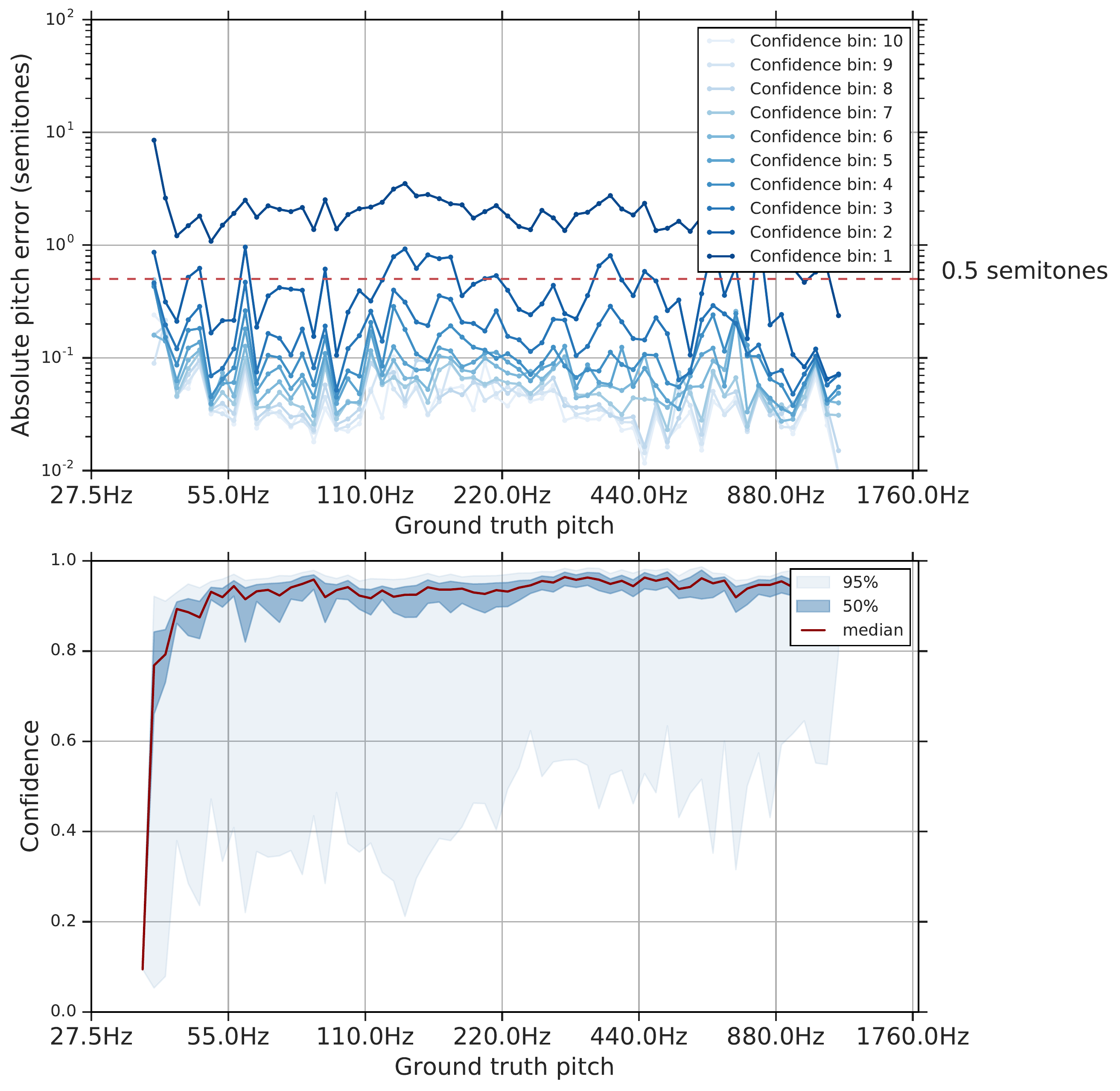}
    \caption{CREPE full}
    \label{fig:error_vs_frequency_crepe_full_mdb_stem_synth}
  \end{subfigure}
  \caption{Pitch error on the {\mdb} dataset, conditional on ground truth pitch and model confidence.}
  \label{fig:pitch_error_mdb}
\end{figure*}

\begin{figure}[t]
  \centering
  \includegraphics[width=0.6\linewidth]{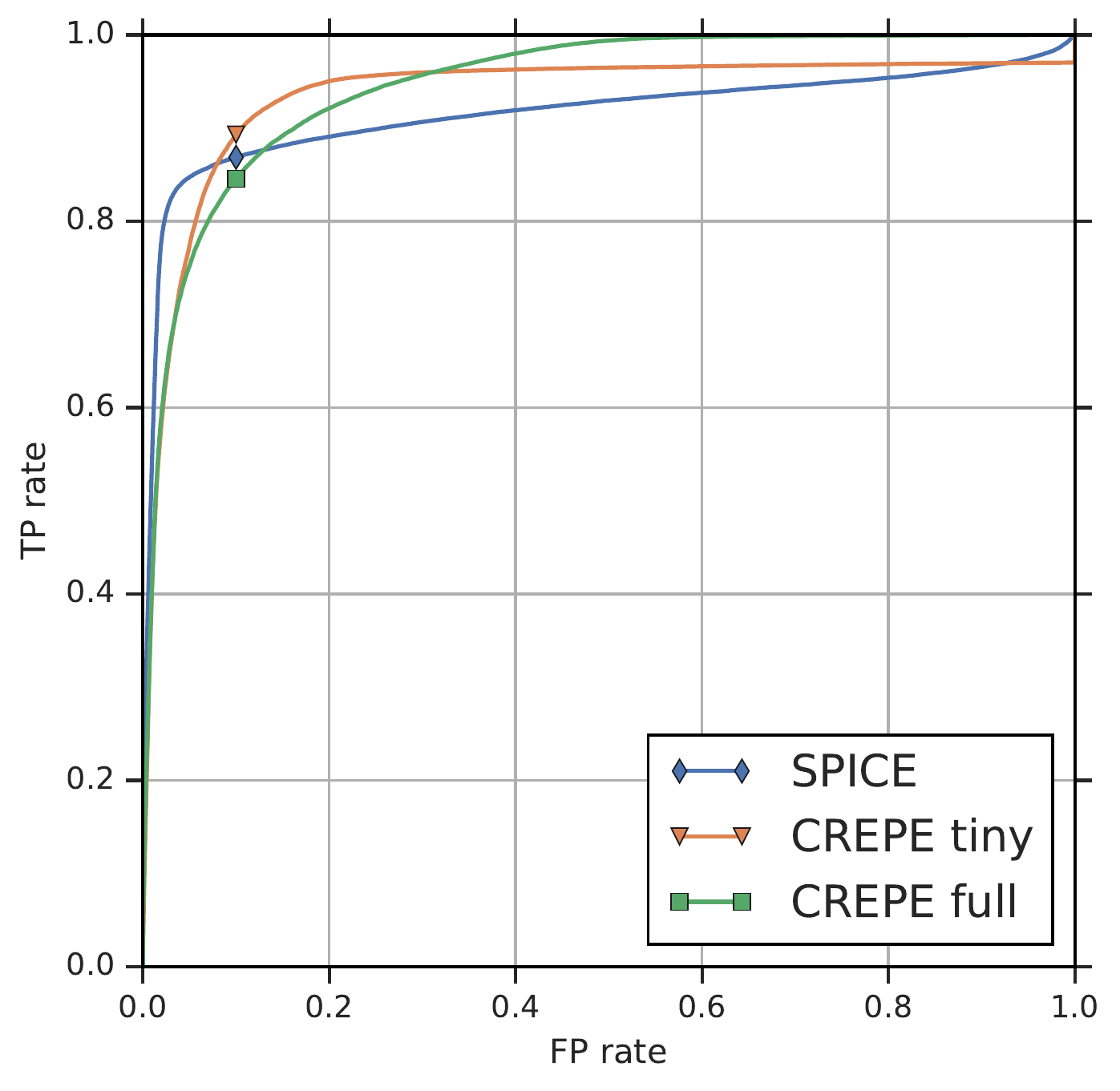}
  \caption{Voicing Detection - ROC ({\mironek}).}
  \label{fig:evaluation_vrr}
\end{figure}

\begin{table*}[t]
  \centering
  \caption{Evaluation results on noisy datasets.}\label{tab:noisy_eval}
   \begin{tabular}{lcccccc}
      \hline
      & & & \multicolumn{4}{c}{{\mironek}} \\
      Model & \# params & Trained on & clean & 20dB &  10dB &  0dB\\

      \hline
      SWIPE & - & - & 86.6\% & 84.3\% & 69.5\% & 27.2\% \\
      CREPE tiny & 487k & many & 90.7\% & 90.6\% & 88.8\% & 76.1\% \\
      CREPE full & 22.2M & many & 90.1\% & 90.4\% & 89.7\% & 80.8\% \\
      SPICE & 2.38M & {\mironek} + augm. & $91.4\% \pm 0.1\%$
      & $91.2\% \pm 0.1\%$ & $90.0\% \pm 0.1\%$ & $81.6\% \pm 0.6\%$ \\
      \hline
   \end{tabular}
\end{table*}

\subsection*{Baselines}
We compare our results against two baselines, namely SWIPE~\cite{Camacho2008}
and CREPE~\cite{Kim2018a}. SWIPE estimates the pitch as the fundamental
frequency of the sawtooth waveform whose spectrum best matches the spectrum of
the input signal. CREPE is a data-driven method which was trained in
a fully-supervised fashion on a mix of different datasets, including
{\mdb}~\cite{Salamon2017}, {\mironek}~\cite{Jang2009},
\emph{Bach10}~\cite{ZhiyaoDuan2010}, \emph{RWC-Synth}~\cite{Mauch2014},
\emph{MedleyDB}~\cite{Bittner2014} and \emph{NSynth}~\cite{Engel2017a}. We
consider two variants of the CREPE model, by using model capacity
\emph{tiny} or \emph{full}, and we disabled Viterbi smoothing, so as to
evaluate the accuracy achieved on individual frames. These models have,
respectively, 487k and 22.2M parameters. CREPE also produces a confidence score
for each input frame.

\subsection*{Evaluation measures}
We use the evaluation measures defined in~\cite{Salamon2014} to evaluate and
compare our model against the baselines. The \emph{raw pitch accuracy (RPA)} is
defined as the percentage of voiced frames for which the pitch error is less
than 0.5 semitones.
To assess the robustness of the model accuracy to the initialization, we also
report the interval $\pm 2 \sigma$. Here $\sigma$ is the sample standard
deviation of the RPA values computed using the last 10
checkpoints of 3 separate replicas.
For CREPE we do not report such interval,
because we simply run the model provided by the CREPE authors on each of the
evaluation datasets.
The \emph{voicing recall rate (VRR)} is the proportion of
voiced frames in the ground truth that are recognized as voiced by the
algorithm. We report the VRR at a target voicing false alarm rate equal to 10\%.
Note that this measure is provided only for {\mironek}, since {\mdb} is a
synthetic dataset and voicing can be determined based on a simple silence
thresholding.

\subsection*{Main results}
The main results of the paper are summarized in Table~\ref{tab:main_eval} and
Figure~\ref{fig:evaluation_rpa}.
On the {\mironek} dataset, SPICE outperforms
SWIPE, while achieving the same accuracy as CREPE in terms of RPA (90.7\%),
despite the fact that it was trained in an unsupervised fashion and
CREPE used {\mironek} as one of the training datasets.
Figure~\ref{fig:pitch_error_mir_onek} illustrates a
finer grained comparison between SPICE and CREPE (full model), measuring the
average absolute pitch error for different values of the ground truth pitch
frequency, conditioned on the level of confidence (expressed in deciles)
produced by the respective algorithm. When excluding the decile with low
confidence, we observe that above 110Hz, SPICE achieves an average error around
0.2-0.3 semitones, while CREPE around 0.1-0.5 semitones.

\begin{figure*}[t]
  \begin{subfigure}{0.49\textwidth}
    \centering
    \includegraphics[width=0.99\linewidth]{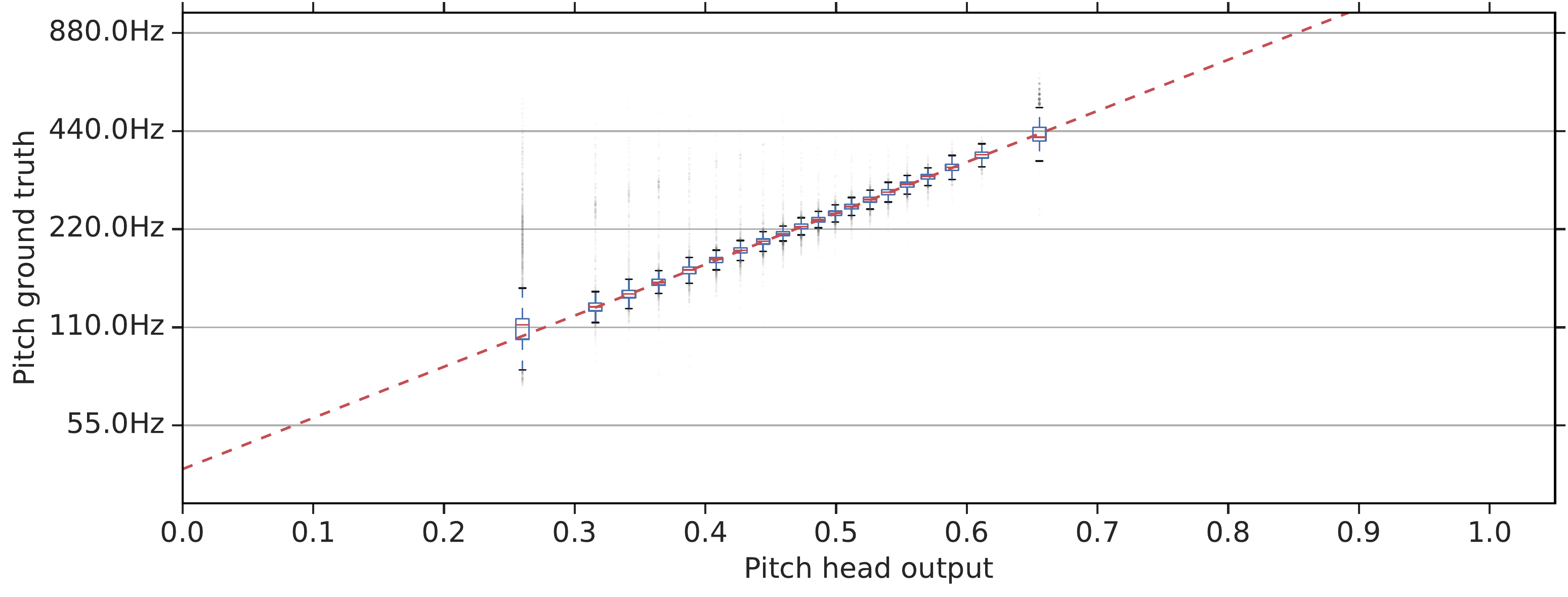}
    \caption{{\mironek}}
    \label{fig:calibration_mironek}
  \end{subfigure}
  \begin{subfigure}{0.49\textwidth}
    \centering
    \includegraphics[width=0.99\linewidth]{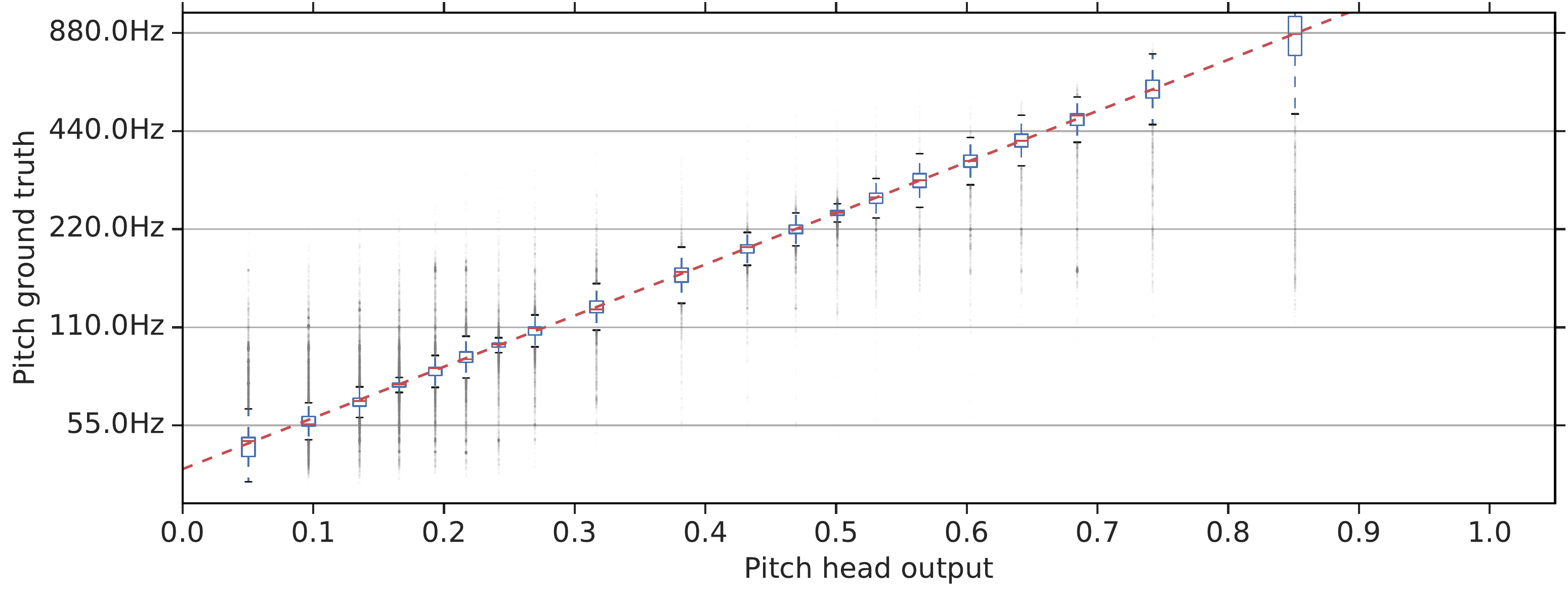}
    \caption{{\mdb}}
    \label{fig:calibration_mdb}
  \end{subfigure}
  \caption{Calibration of the pitch head output.}
  \label{fig:calibration}
\end{figure*}

\begin{figure}[t]
  \centering
  \includegraphics[width=1.00\linewidth]{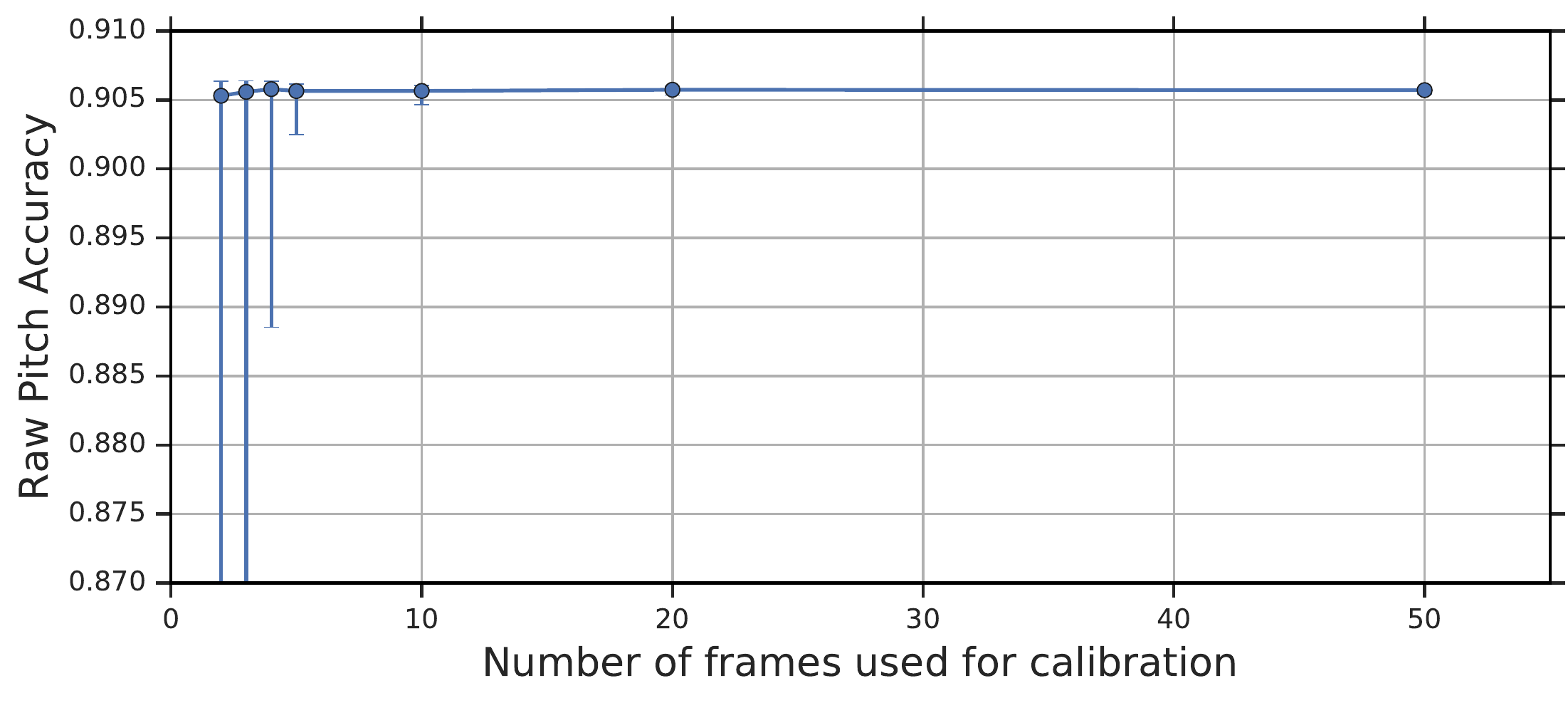}
  \caption{Robustness of the RPA on {\mironek} when varying the number of synthetically generated samples used for calibration.}
  \label{fig:calibration_robustness_mir_onek}
\end{figure}

We repeated our analysis on the {\mdb} dataset. In this case the dataset has
remarkably different characteristics from the {\googlers} dataset used for the
unsupervised training of SPICE, in terms of both frequency extension
(Figure~\ref{fig:dataset_pitch_hist}) and timbre (singing vs. musical
instruments). This explains why in this case the gap between SPICE and CREPE is
wider (88.9\% vs. 93.1\%). Figure~\ref{fig:pitch_error_mdb} repeats the
fine-grained analysis for the {\mdb} dataset, illustrating larger errors at both
ends of the frequency range. We also performed a thorough error analysis, trying
to understand in which cases CREPE and SWIPE outperform SPICE. We discovered
that most of these errors occur in the presence of a harmonic signal, in which
most of the energy is concentrated above the fifth-order harmonics, i.e., in the
case of musical instruments characterized by a spectral timbre considerably
different from the one of singing voice.

We also evaluated the quality of the confidence estimation comparing the
\emph{voicing recall rate (VRR)} of SPICE and CREPE. Results in
Table~\ref{tab:main_eval} show that SPICE achieves results comparable with CREPE
(86.8\%, i.e., between CREPE tiny and CREPE large), while being more accurate in
the more interesting low false-positive rate regime (see
Figure~\ref{fig:evaluation_vrr}).

In order to obtain a smaller, thus faster, variant of the SPICE model, we used
the MorphNet~\cite{Gordon2018} algorithm. Specifically, we added to the
training loss~\eqref{eq:total_loss} a regularizer which constrains the number
of floating point operations (FLOPs), using $\lambda = 10^{-7}$ as
regularization hyper-parameter. MorphNet produces as output a slimmed network
architecture, which has 180k parameters, thus more than 10 times smaller than
the original model. After training this model from scratch, we were still able
to achieve a level of performance on {\mironek} comparable to the larger SPICE
model, as reported in Table~\ref{tab:main_eval}.

Table~\ref{tab:ablation_study} shows the results of the ablation study we
carried out to assess the importance of some of the design choices described in
Section~\ref{sec:method}. These results indicate that the reconstruction loss is
crucial to obtain good results. We believe that this loss acts as a regularizer,
as it enforces inputs with the same pitch but different timbre to have the same
latent values. Pitch shift data augmentation is also important, especially on
{\mdb}, which has a wider pitch range than the training dataset ({\googlers}).
Using continuous pitch shift augmentation instead of discrete octave shifts
gives somewhat worse results, likely due to the artefacts it introduces.
Finally, using Huber loss instead of L2 or L1 loss also gives a significant
gain, which we attribute to the fact that some inputs in the data are actually
unvoiced, and hence it is useful to reduce the impact of these unvoiced examples
on the loss.

Table~\ref{tab:noisy_eval} shows the results obtained when evaluating the models
in the presence of background music. We observe that SPICE is able to achieve
a level of accuracy very similar to CREPE across different values of SNR.

\subsection*{Calibration}\label{subsec:calibration}
The key tenet of SPICE is that is an unsupervised method. However, as discussed
in Section~\ref{sec:method}, the raw output of the pitch head can only represent
relative pitch.
To obtain absolute pitch, the intercept $b$ and the slope $s$ in
\eqref{eq:calibration} need to be estimated based on ground truth
labels, which can be obtained using synthetically generated data without having
access to any labelled dataset as described in Section~\ref{sec:method}.
Figure~\ref{fig:calibration} shows the fitted model for both {\mironek} and
{\mdb} as a dashed red line. In order to quantitatively evaluate the robustness
to the calibration process, we generate harmonic waveforms with $K = 3$
higher-order harmonics, with $N = 11$ frames and $H = 512$ samples. Then, we
vary the number of samples $M \in \{2, 3, 5, 10, 20, 50\}$, repeat the
calibration step 100 times and compute the RPA on the {\mironek} dataset.
Figure~\ref{fig:calibration_robustness_mir_onek} reports the results of this
experiment (error bars represent 2.5\% and 97.5\% quantiles). We observe that
using as few as $M=5$ synthetically generated samples are generally enough to
obtain stable results.

\section{Conclusion}\label{sec:conclusions}
In this paper we propose SPICE, a self-supervised pitch estimation algorithm for
monophonic audio. The SPICE model is trained to recognize relative pitch without
access to labelled data and it can also be used to estimate absolute pitch by
calibrating the model using just a few labelled examples. Our experimental
results show that SPICE is competitive with CREPE, a fully-supervised model that
was recently proposed in the literature, despite having no access to ground
truth labels.
The SPICE model is publicly available as a Tensorflow Hub module at
\url{https://tfhub.dev/google/spice/1}.

\section*{Acknowledgment}
We would like to thank Alexandra Gherghina, Dan Ellis, Dick Lyon and the
anonymous reviewers
for their help with and feedback on this work.

\ifCLASSOPTIONcaptionsoff
  \newpage
\fi



%

\bibliographystyle{IEEEtran}
\bibliography{references}

%








\end{document}